\documentclass{PoS}

\title{Latest results from the KASCADE-Grande data analysis}

\ShortTitle{Latest results from KASCADE-Grande}

\author{D.~ Kang$^{1}$, \speaker{A.~ Haungs$^{1}$},
W.D.~ Apel$^{1}$, J.C.~ Arteaga-Vel\'azquez$^{2}$, K.~ Bekk$^{1}$, M.~ Bertaina$^{3}$,
J.~ Bl\"umer$^{1,4}$, H.~ Bozdog$^{1}$, E.~ Cantoni$^{3,6}$, A.~ Chiavassa$^{3}$,
F.~ Cossavella$^{4}$, K.~ Daumiller$^{1}$, V.~ de Souza$^{7}$, F.~ Di Pierro$^{3}$, P.~ Doll$^{1}$,
R.~ Engel$^{1,4}$, D.~ Fuhrmann$^{8}$, A.~ Gherghel-Lascu$^{5}$, H.J.~ Gils$^{1}$, R.~ Glasstetter$^{8}$,
C.~ Grupen$^{9}$, D.~ Heck$^{1}$, J.R.~ H\"orandel$^{10}$, T.~ Huege$^{1}$,
K.-H.~ Kampert$^{8}$, H.O.~ Klages$^{1}$, K.~ Link$^{4}$, P.~ {\L}uczak$^{11}$, H.J.~ Mathes$^{1}$,
H.J.~ Mayer$^{1}$, J.~ Milke$^{1}$, C.~ Morello$^{6}$, J.~ Oehlschl\"ager$^{1}$,
S.~ Ostapchenko$^{12}$, T.~ Pierog$^{1}$, H.~ Rebel$^{1}$, D.~ Rivera-Rangel$^{2}$, M.~ Roth$^{1}$,
H.~ Schieler$^{1}$, S.~ Schoo$^{1}$, F.G.~ Schr\"oder$^{1}$, O.~ Sima$^{13}$, G.~ Toma$^{5}$,
G.C.~ Trinchero$^{6}$, H.~ Ulrich$^{1}$, A.~ Weindl$^{1}$, J.~ Wochele$^{1}$, J.~ Zabierowski$^{12}$ - KASCADE-Grande Collaboration\footnote{for collaboration list see PoS(ICRC2019)1177}\\
\llap{$^1$} Institut f\"ur Kernphysik, KIT - Karlsruhe Institute of Technology, Germany\\
\llap{$^2$} Universidad Michoacana, Inst.~F\'{\i}sica y Matem\'aticas, Morelia, Mexico\\
\llap{$^3$} Dipartimento di Fisica, Universit\`a degli Studi di Torino, Italy\\
\llap{$^4$} Institut f\"ur Experimentelle Teilchenphysik KIT - Karlsruhe Institute of Technology, Germany\\
\llap{$^5$} Horia Hulubei National Institute of Physics and Nuclear Engineering, Bucharest, Romania\\
\llap{$^6$} Osservatorio Astrofisico di Torino, INAF Torino, Italy\\
\llap{$^7$} Universidade S$\tilde{a}$o Paulo, Instituto de F\'{\i}sica de S\~ao Carlos, Brasil\\
\llap{$^8$} Fachbereich Physik, Universit\"at Wuppertal, Germany\\
\llap{$^9$} Department of Physics, Siegen University, Germany\\
\llap{$^{10}$} Dept. of Astrophysics, Radboud University Nijmegen, The Netherlands\\
\llap{$^{11}$} National Centre for Nuclear Research, Department of Astrophysics, Lodz, Poland\\
\llap{$^{12}$} Frankfurt Institute for Advanced Studies (FIAS), Frankfurt am Main, Germany \\
\llap{$^{13}$} Department of Physics, University of Bucharest, Bucharest, Romania\\
E-mail: \email{donghwa.kang@kit.edu}
}

\abstract{
Over the past 20 years, KASCADE and its extension KASCADE-Grande were dedicated to measure high-energy cosmic rays with primary energies of 100 TeV to 1 EeV. The data accumulation was fully completed and all experimental components were dismantled, though the analysis of the high-quality data is still continued. E.g., we investigated the validity of the hadronic interaction model of the new SIBYLL version 2.3c. We also published a new result of a search for large-scale anisotropies performed with the KASCADE-Grande data. Investigation of the attenuation length of the muon in the atmosphere is also updated with the predictions of the SIBYLL 2.3 interaction model. We investigated, in addition, the muon content of high-energy air showers and compared them to all post-LHC interaction models. In this contribution, the new and updated results from KASCADE-Grande will be presented. An update of the web-based data center KCDC offering the original scientific data from KASCADE-Grande to the public will be briefly discussed as well.
}

\FullConference{36th International Cosmic Ray Conference -ICRC2019-\\
		July 24th - August 1st, 2019\\
		Madison, WI, U.S.A.}

\begin{document}
\setcounter{page}{2}

\makeatletter
\setbox\@firstaubox\hbox{\small D. Kang}
\makeatother

\section{Introduction}
Extensive air shower arrays of KASCADE-Grande \cite{Apel1} and its original array KASCADE \cite{Antoni1}
were located at the Karlsruhe Institute of Technology, Karlsruhe, Germany
(49.1$^{\circ}$ north, 8.4$^{\circ}$ east, 110\ m above sea level),
and devoted to measure high-energy cosmic rays to understand the energy spectrum,
the mass composition, and the arrival direction of cosmic rays.
In particular, the investigations in the energy range of PeV to EeV covered by KASCADE and KASCADE-Grande give important messages to identify the transition region of galactic and extra-galactic cosmic rays.

The data acquisition was completed at the end of 2013 and all experimental components are meanwhile fully dismantled.
A precise analysis of more than 20 years measured data resulted in fruitful findings:
The all-particle energy spectrum reconstructed using the KASCADE data shows
a knee-like structure due to a steepening of spectra of light elements \cite{Antoni2}.
The all-particle energy spectrum measured by KASCADE-Grande \cite{Apel2} shows structures,
which do not follow a single power law:
a concave behavior just above $10^{16}$ eV and a small break at around $10^{17}$ eV,
where a knee-like feature would be expected as the knee of the heavy primaries, mainly iron.
In the reconstructed energy spectra for individual mass groups,
the knee-like feature in the heavy primary spectrum is observed much more significantly at around 80 PeV \cite{Apel3}.
Further, an ankle-like structure is observed at an energy of 100 PeV in the energy spectrum of light primary cosmic rays \cite{Apel4}.
In addition, using full data sets taken by KASCADE and KASCADE-Grande,
we determined upper limits to the flux of ultra-high energy gamma rays, which set constraints on some fundamental astrophysical models \cite{Kang2017}.

The analysis of the measured data of more than 20 years is still in progress.
In this paper, we report on the recently ongoing studies:
the validity of the hadronic interaction model of the new SIBYLL version 2.3c,
a new result of a search for large-scale anisotropies,
and studies of the muon content of extensive air shower based on post-LHC models.
At the end, the status of the KASCADE Cosmic ray Data Centre (KCDC) will be shortly presented as well.

\section{Test of the hadronic interaction model SIBYLL 2.3c}
One of the important analyses after the completeness of measurements
is the test of hadronic interaction models with KASCADE and KASCADE-Grande data.

Recently, a new version of SIBYLL, SIBYLL 2.3c, was released \cite{FelixRiehn}.
It is improved by adjusting particle production spectra
to match the expectation of Feynman scaling in the fragmentation region.
For extensive air showers, the prediction of the updated model is similar to SIBYLL 2.3:
The number of muons is slightly larger than in SIBYLL 2.3.
Compared to SIBYLL 2.1, the number of muons are larger by a factor of about 1.4 at $10^{16}$ eV.

For the air shower simulations the program CORSIKA \cite{Heck}
has been used,
applying different embedded hadronic interaction models.
To determine the signals in the individual detectors, all secondary
particles at the ground level are passed through a detector simulation program
using the GEANT 3.21 package.
The FLUKA (E $<$ 200\ GeV) model has been used for hadronic interactions at low energies.
High-energy interactions were treated with SIBYLL 2.3c, SIBYLL 2.3,
EPOS-LHC, and QGSJetII-04, respectively. 
Showers initiated by five different primaries (H, He, CNO, Si, and Fe nuclei)
have been simulated. The
simulations covered the energy range of 10$^{14}$ - 10$^{18}$ eV with zenith
angles in the interval 0$^{\circ}$ - 42$^{\circ}$. The spectral index in the
simulations was -2 and for the analysis it is converted to a slope of -3. 

By means of the shower size measured by KASCADE-Grande data,
initial tests of SIBYLL 2.3c were performed \cite{Kang1}.

\begin{figure}[t!]
  \begin{center}
    \includegraphics[width=0.52\textwidth]{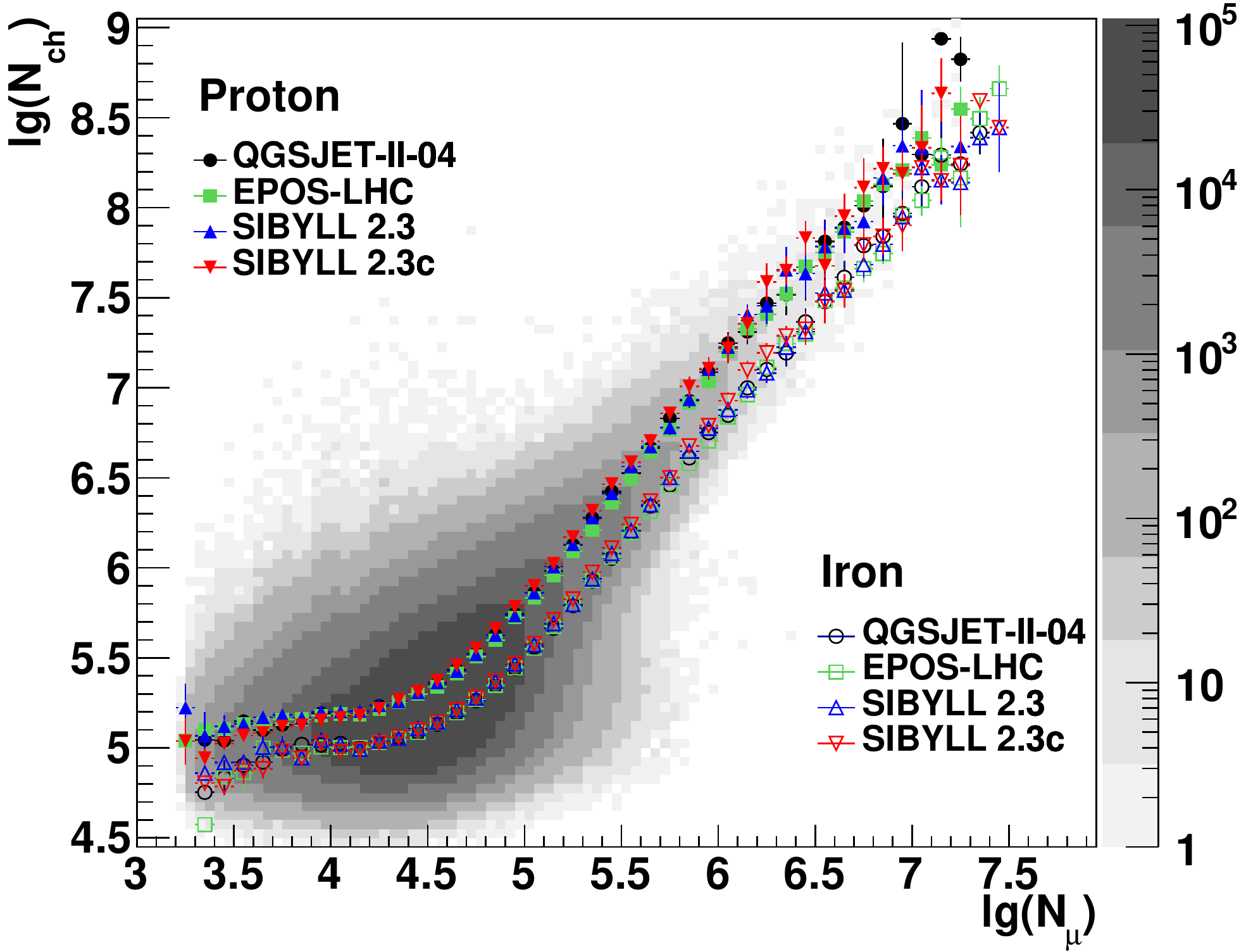}
    \includegraphics[width=0.47\textwidth]{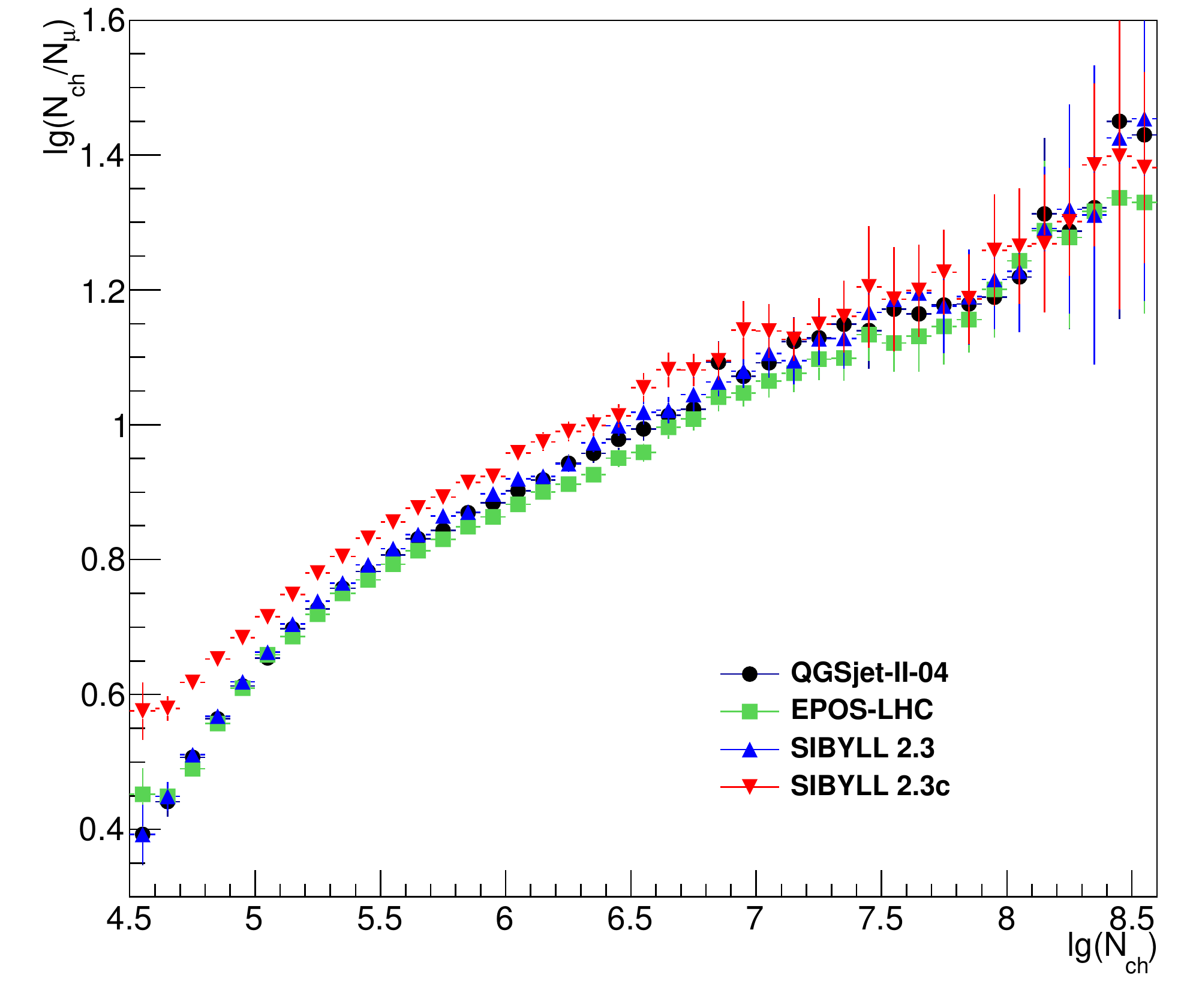}
    \caption{
      {\bf Left:} The 2-dim. shower size distribution measured by KASCADE-Grande,
      along with proton and iron induced showers for different simulations.
      {\bf Right:} The ratio of the total number of charged particles to the total muon numbers.
}
\label{fig1}
\end{center}
\end{figure}

Figure\ 1 (left) shows the shower size measured by KASCADE-Grande, including the full detector response by simulation, along with proton and iron induced showers for the QGSJetII-04, EPOS-LHC, SIBYLL 2.3 and SIBYLL 2.3c simulations.
The solid symbols are for proton induced showers and open ones are for iron, predicted by different interaction models.
SIBYLL 2.3c has a similar tendency to the SIBYLL 2.3 model, but it has less muons compared to EPOS-LHC.

A ratio of the total number of charged particles ($N_{ch}$) to the total muon numbers ($N_{\mu}$)
for different interaction models is shown in the right plot of Fig.\ 1.
Both QGSJetII-04 and SIBYLL 2.3 models have a similar abundance ratio of $N_{ch}$ to $N_{\mu}$, but EPOS-LHC has approximately 10\% more muons, and SIBYLL 2.3c has about 10\% less muons, comparing to QGSJetII-04.
It implies that a less dominant light mass composition is predicted if SIBYLL 2.3c is used to reconstruct the primary mass.

Based on the measured shower size by KASCADE-Grande only,
we reconstructed the primary energy spectrum using 
the energy calibration with the new SIBYLL 2.3c model,
where the atmospheric attenuation effects are corrected
by using the constant intensity cut method.

To reconstruct energy spectra for individual mass groups,
we divided the data into two subsets for heavy and light groups,
based on the $y$ cut method \cite{Kang2}.
The energy calibration function for light and heavy induced showers
is shown in Fig.\ 2 (left).
The slope of SIBYLL 2.3c is quite different from the other two models,
and interestingly two lines get closer together at energies about $10^{18}$eV.
Using this fit function, we converted the attenuation corrected
shower size into the reconstructed energy.
Figure\ 2 (right) presents the reconstructed energy spectra
for light and heavy induced showers.
Interestingly, the spectrum for light primaries gets very close
to the one for heavy primary at energies around 1 EeV.
Compared to other previous SIBYLL models, we see a slight discrepancy
of the spectral slopes due to different ratio of $N_{ch}/N_{\mu}$,
however, all the spectra show a similar feature of the energy spectrum.
The total flux is shifted about 10-20\%, but the general structures are similar.
It is to note that the difference between proton and iron is large for SIBYLL 2.3c.
A detailed estimation of systematic uncertainties is currently being performed. 
It is, however, expected to be about the order of 20\%.

\begin{figure}[]  
\begin{center}
\includegraphics[width=0.45\textwidth]{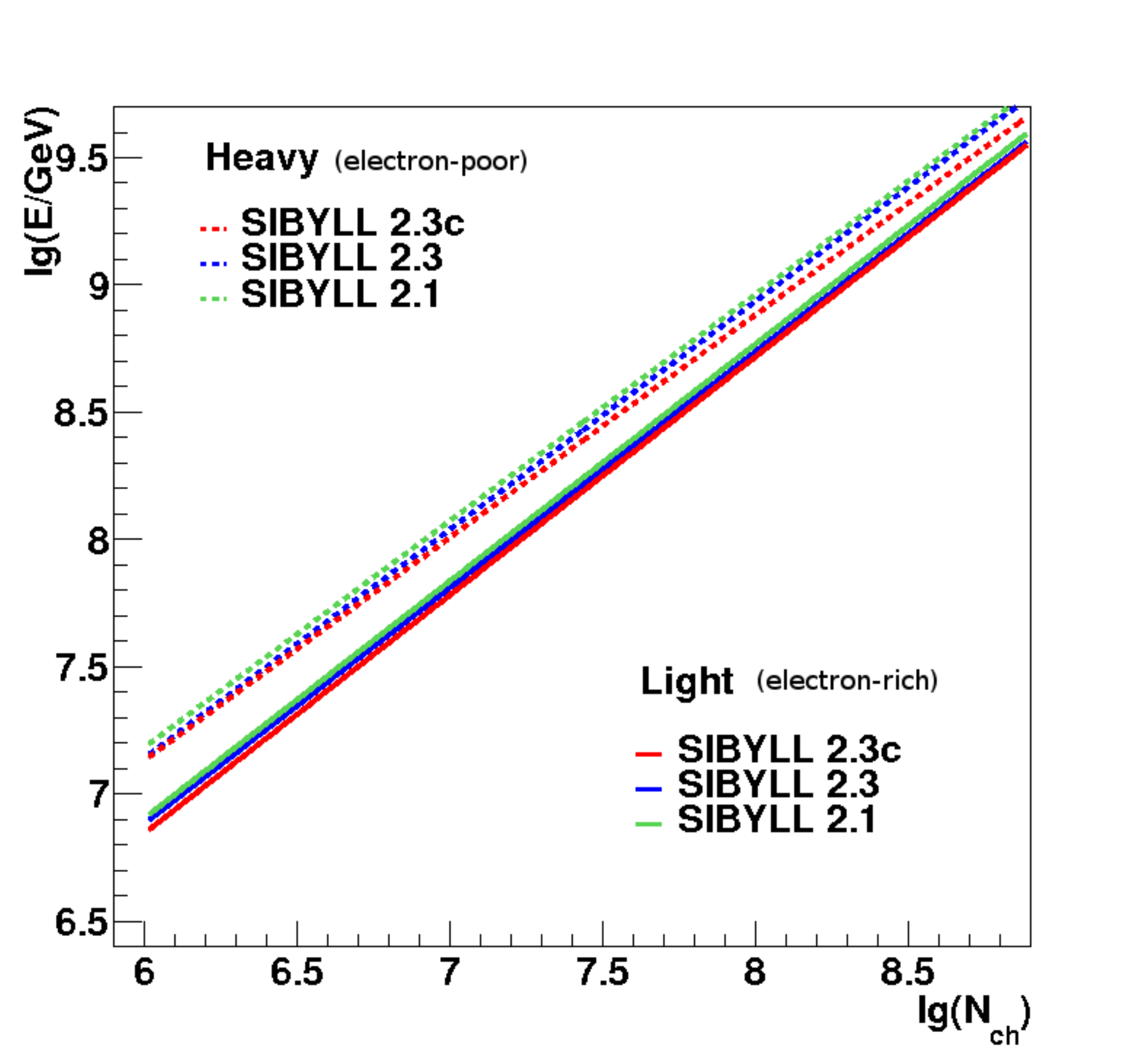}
\includegraphics[width=0.54\textwidth]{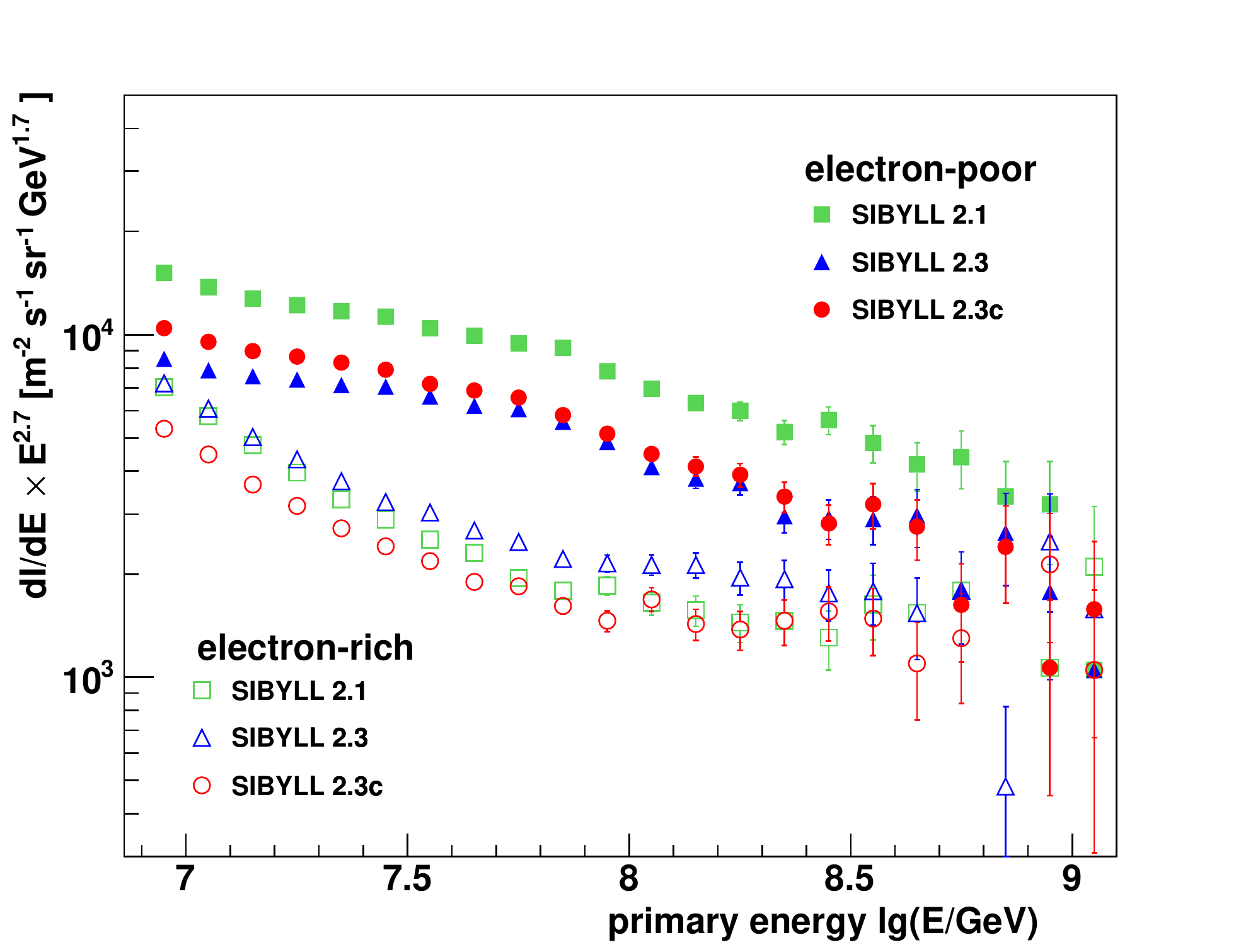}
\caption{
  {\bf Left:} The energy calibration function of light (electron-rich) and
  heavy (electron-poor) primaries for SIBYLL 2.3c, SIBYLL 2.3, and SIBYLL 2.1. 
  {\bf Right:} The resulting energy spectra based on the three different SIBYLL models.
}
\label{fig2}
\end{center}
\end{figure}

\section{Muon content}
In extensive air showers, the particular component of muons
are a sensitive observable for the primary cosmic ray mass
and the physics of hadronic interaction.
However, a precise measurement of the muonic component is difficult in general
due to their low densities in the showers.

As discussed above, the all-particle energy spectra for the different interaction models
show a similar structure, though they 
still do not agree to each other and to data.
This problem might be caused by the muons.
Therefore, we studied the fluctuation
of the muon content of air showers
as a function of primary energy and the zenith angle
by means of KASCADE-Grande data \cite{JuanCarlos}.
We tested, in addition, the predictions of the muon content
for the post-LHC hadronic interaction models:
QGSJetII-04, EPOS-LHC, SIBYLL 2.3 and SIBYLL 2.3c.

\begin{figure}[t!]
  \begin{center}
    \includegraphics[width=0.7\textwidth]{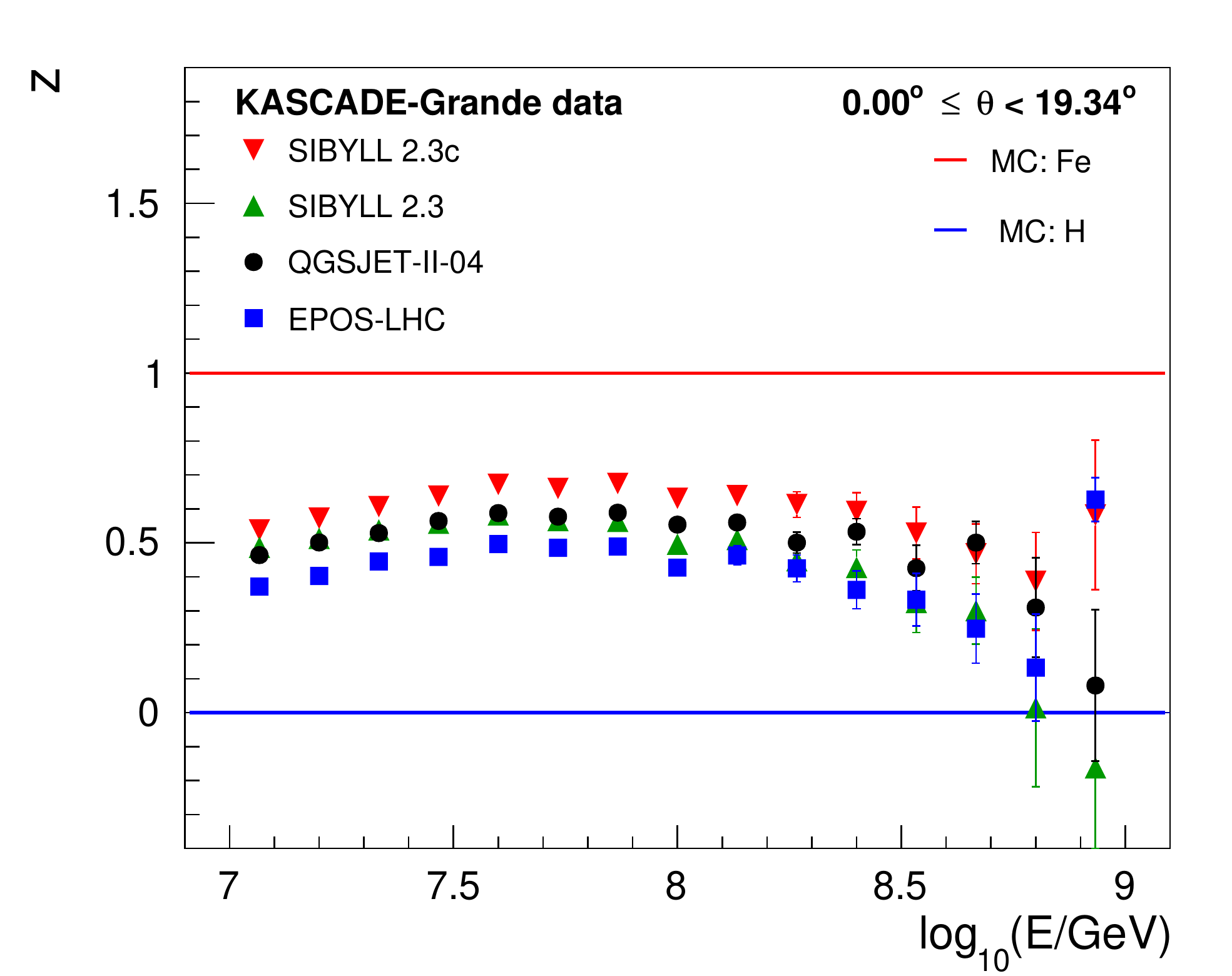}
    \includegraphics[width=0.7\textwidth]{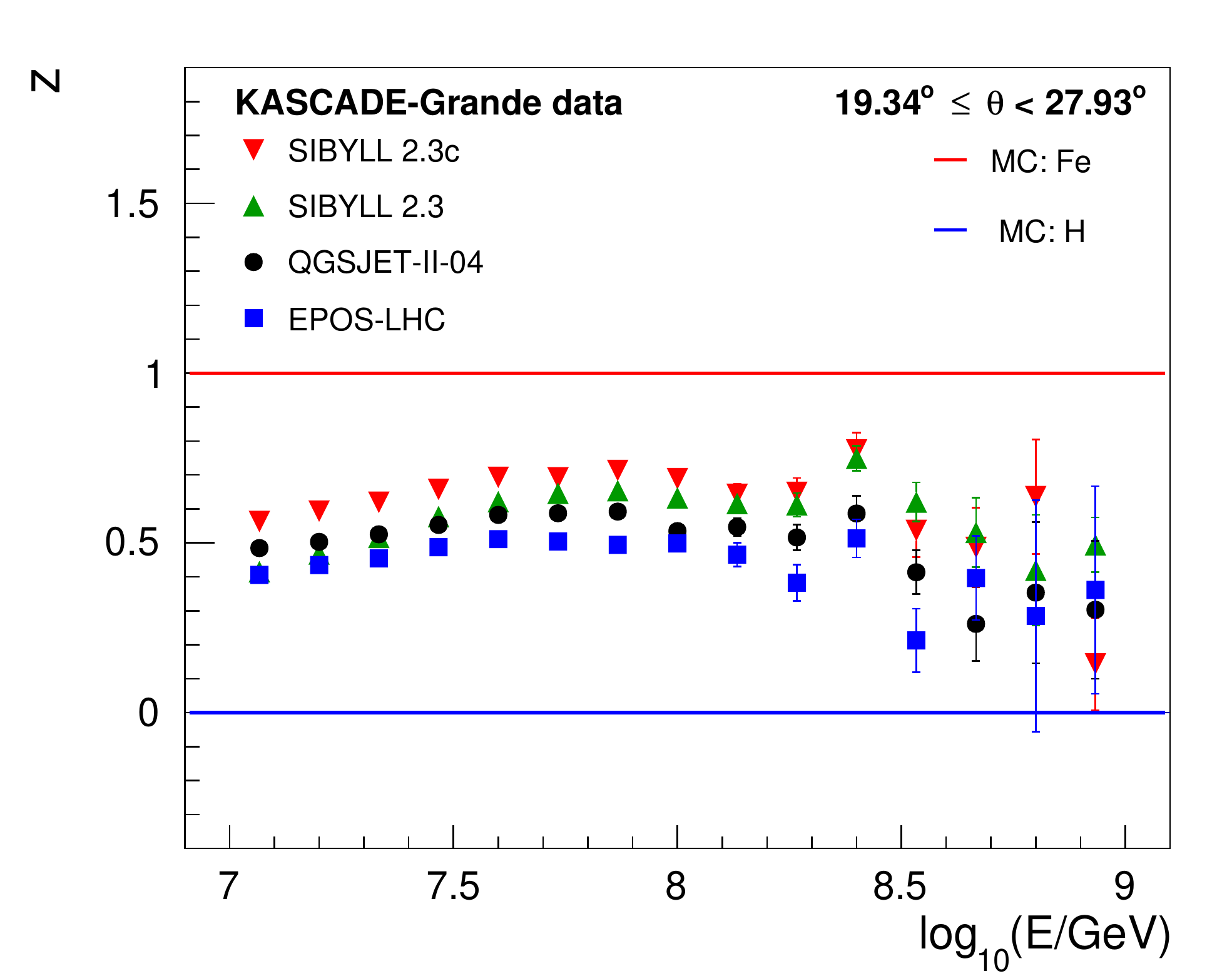}        
    \caption{
      Comparison of measured and predicted mean values
      of the z-scale as a function of the estimated primary energy
      for four different interaction models for two different zenith angle bins.
      The red lines present the expectations for iron and the blue lines for proton \cite{JuanCarlos}.
    }
\label{fig3}
\end{center}
\end{figure}

To compare the muon content of model predictions and measured data, the $z$-scale is used,
which is defined by
\begin{eqnarray}
 z = \frac{ {\rm ln}(N_{\mu}^{det}) - {\rm ln}(N_{\mu}^{p})}{ {\rm ln}(N_{\mu}^{Fe}) - {\rm ln}(N_{\mu}^{p})},
\label{int_eq}
\end{eqnarray}
where $N_{\mu}^{p,Fe}$ are the Monte Carlo predictions
for the muon number induced by proton and iron nuclei.
$N_{\mu}^{det}$ is the measured muon shower size.
A bias between the true and measured muon numbers can be canceled out
in this way of calculating $z$. 
The $z$-scale is defined in such a way that $z$ is 1 for shower muon numbers equal to
the mean of iron nuclei and 0 for $N_{\mu}$ equal to the average of protons.

Figure\ 3 shows the $z$ distribution as a function of the estimated primary energy,
where the energy is estimated event by event
by the standard energy estimation procedure \cite{Apel2}.
The symbols are the experimental data and the lines are predicted mean values
of the $z$ scale for QGSJetII-04, EPOS-LHC, SIBYLL 2.3 and SIBYLL 2.3c.
The total systematic error is not shown in Fig. 3, but
the dominant contribution is from uncertainties in the estimated energy and the shower size,
in particular, the error of the estimated energy becomes larger to higher zenith angles.
Further discussions of the systematic uncertainties are given in Ref. \cite{JuanCarlos}.

The evolution of the mass composition shows a similar behaviour in all cases:
from lighter mean mass at 10 PeV to a heavier mean mass at 100 PeV to lighter again at 1 EeV.
However, the different interaction models predict a different mean mass
and the differences increase with primary energy.
In addition, an inconsistency with zenith angle is visible in the shown
two different zenith angular ranges, where this also increases with higher energy.
There are also other points, like e.g. the evolution of the muon number depending on energy,
where we observed discrepancies between the measured and the simulated
muon attenuation length \cite{JuanCarlos2}.

These muon studies are foreseen to be addressed in the working group report
on the combined analysis of muon density measurements.

\section{Large-scale anisotropy}

\begin{figure}[t!]
  \begin{center}
    \includegraphics[width=0.48\textwidth]{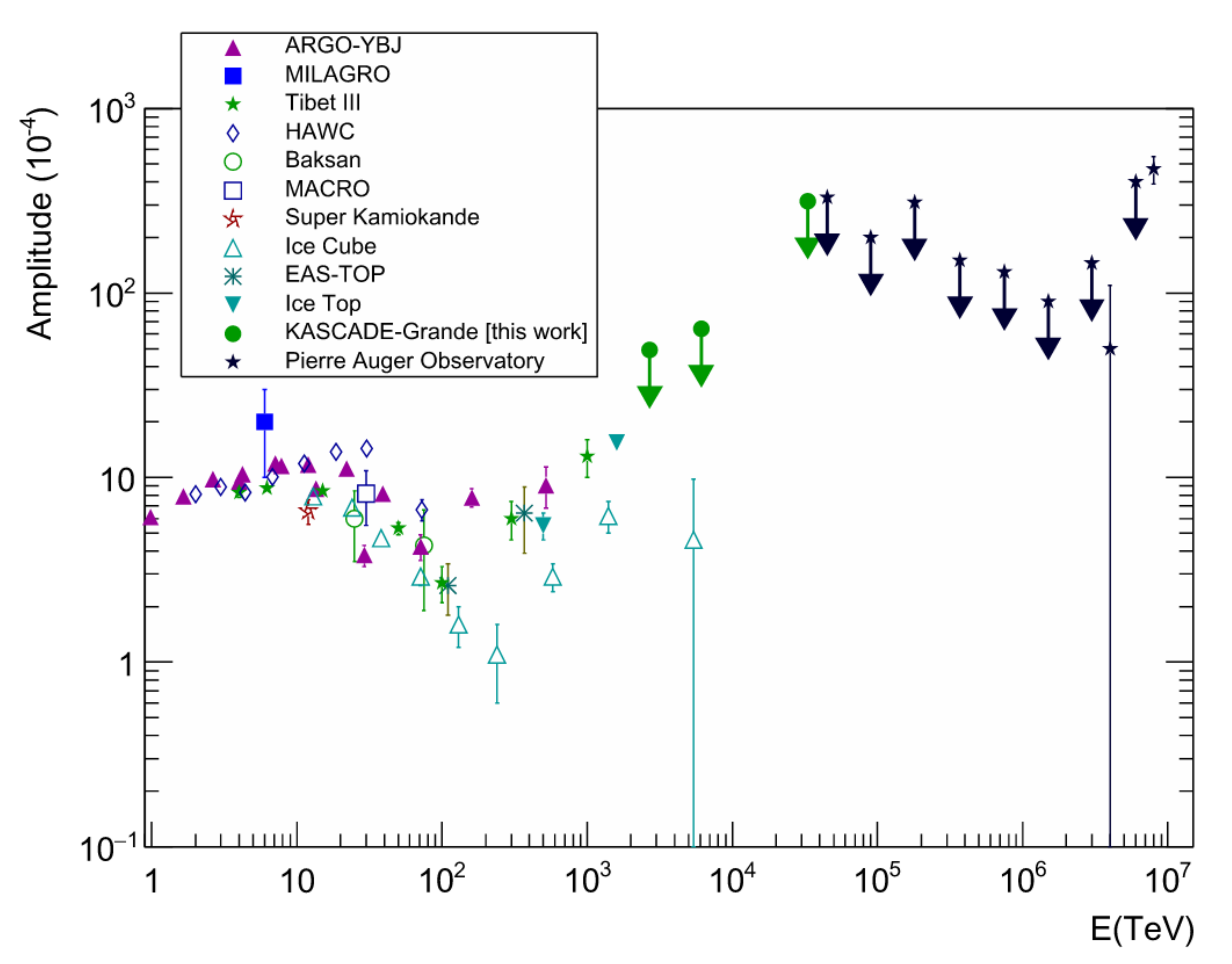}
    \includegraphics[width=0.5\textwidth]{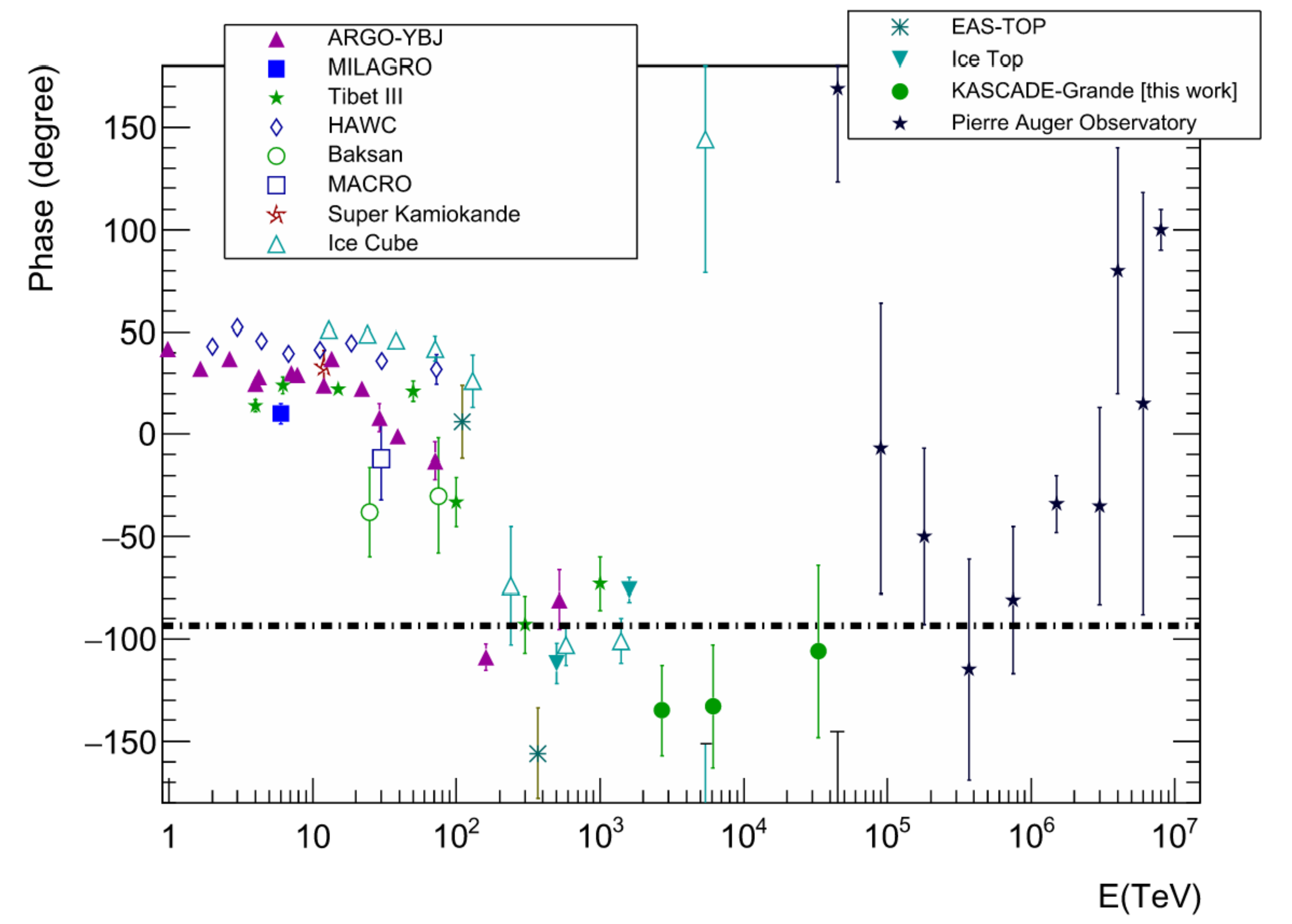}
    \caption{
      Comparison of the KASCADE-Grande measurements of the phase (right)
      and of the upper limits to the amplidude (left)
      of the first harmonic with experimental results
      obtained in the energy range of $10^{13}$ up to $10^{19}$ eV \cite{Apel5}.
      The dashed line shows the direction of the glactic center.
    }
\label{fig4}
\end{center}
\end{figure}
Recently, we presented the results of the search for large-scale anisotropies in the
arrival directions of cosmic rays with the KG data at energies higher than $10^{15}$ eV \cite{Apel5}.

For this analysis, we used the KASCADE-Grande data from December 2003 to October 2011,
which are in total $10^{7}$ events.
The number of counts from the eastward and westward directions are affected by a trigger
inefficiency in the same way, so that the east-west method can also be applied
to the data collected with trigger conditions that do not reach a 100\% efficiency.
No selection cuts on the core position are applied,
and all the events with zenith angle of less than 40$^{\circ}$ are used. 

We used the east-west method in order to eliminate
spurious anisotropies due to atmospheric and instrumental effects.
So, this techniques allow us to remove correctly the count
variations, which are not associated to real anisotropy.

Applying the east-west method, we obtained the number of
counts distributions with 20 min bin width, corresponding to an
angular aperture of 5 degrees in solar, sidereal, and anti-sidereal time.
Then we fitted these distributions with a first harmonic function,
and obtained amplitude and phase,
where the background fluctuation is calculated with Rayleigh probability.
The significance of the amplitude of the first harmonic is 3.5 sigma,
therefore, we calculated an upper limit to the amplitude ($A < 0.47 \times 10^{-2}$)
at the 99\% confidence level.

Figure 4 shows the comparison of the phase (right) and amplitude (left)
measured by KASCADE-Grande
with other experimental results in the energy ranges of $10^{13}$ up to $10^{19}$ eV.
The phase of the first harmonic measured by KASCADE-Grande observed
a change of the phase of the first harmonic
in the direction of the galactic center
at energies $~2 \times 10^{14}$ eV
and it remains at energies greater than the knee of the cosmic-ray spectrum.
This shows the agreement with the other measurements by ESA-TOP, IceCube, and IceTop
at energies higher than $2 \times 10^{14}$ eV.

These results fill the energy range between knee and ankle of the cosmic-ray spectrum.
The phase of the first harmonic changes from about 30$^{\circ}$ to -140$^{\circ}$
at energies above $10^{14}$ eV, then it remains flat until $10^{17}$
and changes again to -100$^{\circ}$ around $8 \times 10^{18}$.
The first phase change of the dipole anisotropy might imply that
the galactic cosmic-ray sources are densely distributed in the galactic center region.
The second phase change could be an indication as a sign of an extra-galactic origin
of ultra high-energy cosmic rays.

\section{KASCADE Cosmic ray Data Centre (KCDC)}
KCDC is a web portal (https://kcdc.ikp.kit.edu),
where data of the KASCADE and KASCADE-Grande experiments are made available
for the interested general public \cite{KCDC2018}.
Since the first release in 2013, KCDC provides to the public users the measured and reconstructed
parameters of air showers. In addition, KCDC provides the conceptual design, how the data can be
treated and processed so that they are also usable outside the community of experts in the research
field. Detailed educational examples make a use also possible for school students and early
stage researchers. The aim of the project KCDC is the installation and establishment of a public
data centre for high-energy astroparticle physics based on the data of the KASCADE experiment.
Moreover, with KCDC we provide to the public the selected data via a custom-made web page.

In the new release, named NABOO in 2017, data from the KASCADE-Grande detector component have been included
to cover a larger part of the energy spectrum. 4.3$\cdot10^{8}$ air shower events are available.
For deeper investigations of the air-shower parameters, e.g. for composition analyses,
full simulations of individual events are necessary.
Thus we published also the full air-shower simulations with the inclusion of the detector responses.
In addition, the data points of nearly 100 energy spectra from many different experiments were published as well.

For the future, the publication of the accompanying software tools for open access will be achieved.
Another plan for the future is to open KCDC for another type of shower data.
Radio data from the LOPES experiment,
which was co-located with KASCADE, will be included.
Due to the different observation technique, the data structure from the LOPES antennas as well as calibration
procedures are different from the ground-based KASCADE experiment, as well as the entire data analysis.
Hence, an adoption of the data platform is required in direction of further generalization of KCDC \cite{Haungs}.

\section{Conclusion}

A validity of the most recent hadronic interaction model of SIBYLL 2.3c is tested, based on the shower size measurements by KASCADE-Grande.
The total energy flux is shifted by roughly 10\%. All structures of energy spectra are similar, however, relative abundancies depend strongly on high-energy hadronic interaction models. 
Predictions of interaction models still do not agree to each other and to data due to muons, therefore, we studied the muon content of air showers through the $z$-value, based on the post-LHC models. An inconsistency with the zenith angle was seen, in particular, at higher primary energies.

Using full data sets taken by KASCADE-Grande, the results of the search for large-scale anisotropies in the arrival directions of cosmic rays at energies higher than $10^{15}$ eV are presented.
The result from this investigation supports the hypothesis of a change of the phase of the first harmonic at energies greater than $~2 \times 10^{14}$ eV.

KCDC is a pioneering work in public access of astroparticle physics data and is already accepted by the astroparticle physics community.
Since astroparticle physics experiments are globally distributed and the community requests for multi-messenger analyses,
further steps of KCDC towards a global data and analysis centre for astroparticle physics
are planned.

\end{document}